\documentstyle[12pt]{article}
\newcommand{\beq}{\begin{equation}}
\newcommand{\eeq}{\end{equation}}
\newcommand{\nin}{\noindent}

\newcommand{\bunit}{\mbox{\bf {1}}}
\newcommand{\bu}{\mbox{\bf {U}}}
\newcommand{\br}{\mbox{\bf {R}}}
\newcommand{\bz}{\mbox{\bf {Z}}}
\catcode`\@=11
\long\def\@makefntext#1{
\protect\noindent \hbox to 3.2pt {\hskip-.9pt
$^{{\ninerm\@thefnmark}}$\hfil}#1\hfill}                

 \def\@makefnmark{\hbox to 0pt{$^{\@thefnmark}$\hss}}  

\def\ps@myheadings{\let\@mkboth\@gobbletwo
\def\@oddhead{\hbox{}
\rightmark\hfil\ninerm\thepage}
\def\@oddfoot{}\def\@evenhead{\ninerm\thepage\hfil
\leftmark\hbox{}}\def\@evenfoot{}
\def\sectionmark##1{}\def\subsectionmark##1{}}

\newcounter{sectionc}\newcounter{subsectionc}\newcounter{subsubsectionc}
\renewcommand{\section}[1] {\vspace{0.6cm}\addtocounter{sectionc}{1}
\setcounter{subsectionc}{0}\setcounter{subsubsectionc}{0}\noindent
	{\bf\thesectionc. #1}\par\vspace{0.4cm}}
\renewcommand{\subsection}[1] {\vspace{0.6cm}\addtocounter{subsectionc}{1}
	\setcounter{subsubsectionc}{0}\noindent
	{\it\thesectionc.\thesubsectionc. #1}\par\vspace{0.4cm}}
\renewcommand{\subsubsection}[1]
{\vspace{0.6cm}\addtocounter{subsubsectionc}{1}
	\noindent {\rm\thesectionc.\thesubsectionc.\thesubsubsectionc.
	#1}\par\vspace{0.4cm}}

\newcounter{appendixc}
\newcounter{subappendixc}[appendixc]
\newcounter{subsubappendixc}[subappendixc]

\renewcommand{\appendix}[1] {\vspace{0.6cm}
	\refstepcounter{appendixc}
	\setcounter{figure}{0}
	\setcounter{table}{0}
	\setcounter{equation}{0}
	\renewcommand{\thefigure}{\Alph{appendixc}.\arabic{figure}}
	\renewcommand{\thetable}{\Alph{appendixc}.\arabic{table}}
	\renewcommand{\theappendixc}{\Alph{appendixc}}
	\renewcommand{\theequation}{\Alph{appendixc}.\arabic{equation}}
	\noindent{\bf Appendix \theappendixc #1}\par\vspace{0.4cm}}

\def\abstracts#1{{
	\centering{\begin{minipage}{30pc}\tenrm\baselineskip=12pt\noindent
	\centerline{\tenrm ABSTRACT}\vspace{0.3cm}
	\parindent=0pt #1
	\end{minipage}}\par}}

\renewenvironment{thebibliography}[1]
	{\begin{list}{\arabic{enumi}.}
	{\usecounter{enumi}\setlength{\parsep}{0pt}
\setlength{\leftmargin 1.25cm}{\rightmargin 0pt}
	 \setlength{\itemsep}{0pt} \settowidth
	{\labelwidth}{#1.}\sloppy}}{\end{list}}

\newcounter{itemlistc}
\newcounter{romanlistc}
\newcounter{alphlistc}
\newcounter{arabiclistc}

\newcommand{\fcaption}[1]{
	\refstepcounter{figure}
	\setbox\@tempboxa = \hbox{\tenrm Fig.~\thefigure. #1}
	\ifdim \wd\@tempboxa > 6in
	   {\begin{center}
	\parbox{6in}{\tenrm\baselineskip=12pt Fig.~\thefigure. #1}
	    \end{center}}
	\else
	     {\begin{center}
	     {\tenrm Fig.~\thefigure. #1}
	      \end{center}}
	\fi}

\newcommand{\tcaption}[1]{
	\refstepcounter{table}
	\setbox\@tempboxa = \hbox{\tenrm Table~\thetable. #1}
	\ifdim \wd\@tempboxa > 6in
	   {\begin{center}
	\parbox{6in}{\tenrm\baselineskip=12pt Table~\thetable. #1}
	    \end{center}}
	\else
	     {\begin{center}
	     {\tenrm Table~\thetable. #1}
	      \end{center}}
	\fi}

\def\@citex[#1]#2{\if@filesw\immediate\write\@auxout
	{\string\citation{#2}}\fi
\def\@citea{}\@cite{\@for\@citeb:=#2\do
	{\@citea\def\@citea{,}\@ifundefined
	{b@\@citeb}{{\bf ?}\@warning
	{Citation `\@citeb' on page \thepage \space undefined}}
	{\csname b@\@citeb\endcsname}}}{#1}}

\newif\if@cghi
\def\cite{\@cghitrue\@ifnextchar [{\@tempswatrue
	\@citex}{\@tempswafalse\@citex[]}}
\def\citelow{\@cghifalse\@ifnextchar [{\@tempswatrue
	\@citex}{\@tempswafalse\@citex[]}}
\def\@cite#1#2{{$\null^{#1}$\if@tempswa\typeout
	{IJCGA warning: optional citation argument
	ignored: `#2'} \fi}}

\def\fnt#1#2{\footnotetext{\kern-.3em
	{$^{\mbox{\sevenrm #1}}$}{#2}}}

 1
 1
 1

\font\tenbf=cmbx10
\font\tenrm=cmr10
\font\tenit=cmti10

\font\ninerm=cmr9

\renewcommand{\thepage}{\arabic{page}}

\begin{document}
NBI-95-15\\
GUTPA/95/04/1\\

\centerline{\tenbf NONLOCALITY AS AN EXPLANATION FOR FINETUNING}
\baselineskip=16pt
\centerline{\tenbf AND FIELD REPLICATION IN NATURE
\footnote{presented by H. B. Nielsen at The 7th Adriatic Conference
``Theoretical and Experimental Perspectives in Particle Phenomenology.
Island of Brioni, 13-20 September 1994.}}
\vspace{0.8cm}
\centerline{\tenrm D. L. BENNETT}
\baselineskip=13pt
\centerline{\tenit Royal Danish College of Pharmacy \&
Niels Bohr Institute, Copenhagen}
\vspace{0.3cm}
\centerline{\tenrm C. D. FROGGATT}
\baselineskip=13pt
\centerline{\tenit Department of Physics and Astronomy, University of Glasgow}
\vspace{0.3cm}
\centerline{\tenrm and}
\vspace{0.3cm}
\centerline{\tenrm H. B. NIELSEN}
\baselineskip=13pt
\centerline{\tenit Niels Bohr Institute, Copenhagen}
\vspace{0.9cm}

\abstracts{Constants of Nature that have nongeneric values pose a riddle
often referred to as the finetuning problem.
The conspicuous values assumed by many physical
constants (e.g., the vanishing effective  cosmological constant, the smallness
of the Higgs mass compared to the Planck scale, the finestructure constants,
$\Theta_{QCD}$) seem to coincide with
values that are obtained if one assumes that
Nature in general seeks out multiple
point values for intensive parameters.
Multiple point values would occur
in the presence of many coexisting phases.
Such coexistence could be enforced by having
fixed but not finetuned amounts of extensive quantities. We show that
universally fixed amounts of extensive quantities is tantamount to having
long range nonlocal interactions of a special type: these interactions
are identical between fields at all pairs of spacetime points regardless
of the spacetime distance between them.
Such omnipresent nonlocal interactions, which can be
described by a very general form of a reparameterization invariant action,
would not be perceived as ``action at a distance'' but rather most likely
incorporated into our theory as constants of Nature. Hence one can
speculate that this mild form
of nonlocality is the underlying explanation of Nature's affinity for the
multiple point. We also speculate that nonlocal effects, described by
fields depending on two spacetime points, may be responsible for the
replication of the fields in three generations. Such a nonlocal
mechanism would also triple the number of boson fields, as in the
antigrand unification model. We briefly review the multiple point
predictions for the three fine structure constants and the
resolution of the quark-lepton mass hierarchy problem in this antigrand
unified extension of the Standard Model.}

\vspace{0.8cm}
\rm\baselineskip=14pt

\newpage

\section{Introduction}
One has a finetuning problem when the experimental values of physical
constants are found to have very special values relative to an a priori
expectation.
An explanation of
why constants of Nature have seemingly nongeneric values cries out for
a theoretical explanation.
Why for instance is the cosmological
constant so exceedingly small in terms of Planck scale units, which one
would naturally suspect were the fundamental units in Nature?
Why is the Higgs expectation value, which determines the
weak interaction scale, so small compared to
the Planck mass or,
if one believes in Grand Unification,
to the unification scale? Addressing the finetuning problems offers
the hope of being able to use hints coming directly from Nature -
rather than from
pure speculation - to learn about what the physics at much shorter distances
than those presently accessible and known should be like.

In this spirit we should like to present some relations involving the
finestructure
constants\cite{pred,picek,glas,alb,van,long,lap,gosen,mit,ngen} -
on which we have worked for
a rather long time - as being finetuning problems. This is meant in the
sense that the finestructure constants rather remarkably
take the values at ``the'' multiple point. In the formulation that we have
used so far, the multiple point is the point in the
phase diagram of a lattice gauge theory (having a sufficiently
general plaquette action)  at which all - or at least many -
phases convene.
Actually the experimental values of the finestructure constants only
coincide with the multiple point values if we make the
assumption of
an ``anti-grand unification model'' \cite{pred,van,frogniel,confusion,gaeta}
with the gauge group $SMG^3$ (which is the 3-fold cartesian product of the
``usual'' Standard Model Group ($SMG$): $SMG^3\stackrel{def}{=}
SMG\times SMG\times SMG$).
The usual Standard Model Group
$SMG=S(U(2)\times U(3))$ arises as the group surviving the Planck scale
breakdown of $SMG^3$ to its diagonal subgroup.
With the anti-grand unified gauge group $SMG^3$,
each generation acquires its own 12 gauge fields
just as in the Standard Model).
Hence, if our gauge group $SMG^3$ etc. is accepted, it is indeed
a finetuning problem
that is addressed in explaining why the finestructure constants
should take just the multiple point
values, on a  par with explaining the smallness of for example the
cosmological constant.

In fact, we would actually suggest
that all the finetuning problems, including the finestructure
constant one, are unified (or at least reformulated)
if it is assumed that Nature in general has an affinity for
the multiple point, where
a lot of phases meet for a single combination of the
``intensive parameters''. The latter are really just parameters of the action.
Included among such parameters -
generalized ``coupling constants'' - are lattice artifact parameters.
This is because we take the lattice as really existing, in the sense that
a lattice is one of many ways of implementing what we assume
to be the actual existence of a
fundamental regularization at roughly the Planck scale.
This assumption is inspired by
the fact that the consistency of any field theory seems to
require a cutoff.

So our basic explanation for the finetuning problems
is that, for some reason, the coupling constants etc. in Nature
take values that correspond to the multiple point where
``all'' (or as many as possible)  phases convene.

An analogous phenomenon is known from other fields of physics:
e.g., a mixture of ice and water (and vapour)
chooses its temperature and pressure to be that of the
melting point (the triple point). By mixtures of well chosen
but not finetuned amounts of various different molecules, it would be possible
to realize a multiple point with more than just three phases that
convene.
Here it is the enforced coexistence (i.e., by insuring that extensive
quantities such as mole number, energy and volume lie within a range
that enforces such coexistence)
of the phases that consequently enforces the
multiple point values for the chemical potentials, temperature
and pressure so that there is a balance w.r.t. exchange
of molecules between the phases.

It is very tempting to speculate that an analogous scenario, in
which there are (e.g., primordially) fixed but not finetuned amounts
of perhaps a
great many extensive quantities (analogous to number of molecules, energy and
volume in the above example),  can provide
an explanation for all finetuned quantities in Nature. We shall see in the
sequel that having fixed amounts of extensive quantities in, say, the universe
implies a mild form of nonlocality (or vice versa)
that, in turn, implies multiple point
criticality and thereby universally fixed physical constants.


\section{The need for nonlocality}

In this section it is argued that at least the cosmological
constant finetuning problem really calls for the breakdown
of the principle of locality in the mild sense referred to above.
Any finetuning problem
concerning coupling
constants - among which we may also include the cosmological constant -
calls for some way by which
                   these coupling constants are rendered ``dynamical'', in
the sense that their values are not simply fixed a priori
but can in some way take on values that must though (for the sake of
translational invariance) be maintained at a constant
value. That a physical constant (e.g., coupling constant) can depend
on something (i.e., in spite of being constant
as a function of spacetime, is not simply fixed a priori) is
the most important content of the baby-universe\cite{baby} theory. The latter
theory
indeed aspires with some success to solve the cosmological constant problem.
The baby-universe theory also makes use of an effective
breakdown of the principle of locality and renders the coupling
constants dynamical. Hereby this theory has the right
ingredients needed
if the goal of explaining why the cosmological
constant is small is to be achieved \footnote{Tsamis and Woodard\cite{woodard}
may have a way around this.}.

The problem in local theories - i.e., healthy theories inasmuch as
locality is seemingly well confirmed - is that, if the coupling
constants and in particular the (bare) cosmological constant are
``dynamical'', the strict validity of a  principle of
locality in
the theory would imply that the bare dynamical cosmological
constant could only depend on the situation at the space time point
in question and, indirectly, also on previous times but certainly not
on the future!  However, a bare cosmological constant that is constant
in spacetime should already in the first moment after the
Big Bang have had its value finetuned once and for all - up to, say, 120
decimal
places -
to the value which makes the {\em dressed} (renormalized) cosmological
constant so exceedingly small (as only can be seen in a background
so depleted of matter as is the case today). That means that the bare constant
had to ``know''
about the details of a vacuum that did not exist at the time when
the bare cosmological constant was already tuned in to the vacuum that would
eventually evolve! Such a tuning of the bare cosmological constant
seems to need some form of
precognition! But this is precisely what is achieved by
breaking the principle of locality.
So we are forced to accept that at least a strict principle of locality
is not allowed,
if we are to explain  the cosmological constant problem in a way
commensurate with having
dynamical (bare) couplings and the renormalization
corrections of quantum field theory with a well defined vacuum.

\section{Retrieving locality}

In Ref. 11 it is argued that, even
if the principle of locality were indeed broken at the
fundamental level, one could imagine how locality might be regained
effectively by restricting the breakdown of locality to a form that does not
violate the principle of general reparameterization invariance
of general relativity.

A theory having nonlocalities extending only over
fundamental scale distances may usually be considered local
when viewed at distances long compared to the fundamental scale.
So the form of nonlocality that potentially could be in conflict with
the phenomenologically  obeyed
principle of locality must involve distances much longer than
the fundamental scale (the Planck scale say). We want to argue that
even nonlocality over extremely large distances
is not in conflict with what we
regard as the phenomenological validity of the principle of locality,
{\bf \em if} the (long distance) nonlocality is restricted to
being invariant under diffeomorphisms or reparameterization.
This class of nonlocality includes that of interest to us - namely
nonlocal interactions surviving
at distances much longer than the fundamental scale {\bf and}
that {\em are the same between the fields at any
pair of points in spacetime independent of the distance between
these points}.

It can be argued\cite{frogniel} that quantum gravity fluctuations
will at large distances $\int_x^yds$ smooth out  the effective interaction
between a pair of fields $\phi(x)$ and $\phi(y)$,
in such a way that interaction coefficients $c(x,y)$ decay exponentially
as a function of distance to values independent of
the distance $\int_x^yds$: i.e., $c(x,y)=const.$ Here the $c(x,y)$ are
defined by there being an action term

\begin{equation}
\int\int d^4xd^4y\sqrt{g(x)g(y)}c(x,y){\cal L}_i(x){\cal L}_j(y) \label{eq1}.
\end{equation}

The expected exponential decay of $c(x,y)$ to the long distance
constant value $const.$ has decay rates not
differing by more than a few orders of magnitude from the fundamental scale.
Hence, for the purposes of very long distances, Eq.~\ref{eq1} becomes
$$ const.\cdot\int\int d^4xd^4y \sqrt{g(x)}\sqrt{g(y)}
{\cal L}(x)_i{\cal L}(y)_j\stackrel{def}{=}const.\cdot I_iI_j$$
The interaction between a number of fields can similarly be taken into
consideration, in such a way that the long distance physics takes the form
of nonlinear functions of integrals $I_j = \int d^4x\sqrt{g(x)} {\cal L}_j(x)$.
Here the ${\cal  L}_i(x)$'s denote expressions of the type that could be
usual Lagrangian density terms.
The reparameterization invariance
of general relativity is in essence assumed in this argumentation.

Indeed  a
principle like reparameterization invariance
is needed, in order to have a symmetry between all
pairs of spacetime points that implies
the same interaction between all such pairs regardless of the distance
separating them.

The important point is that an interaction that has the character of
being the same between the fields located at {\em any} pair of points
(regardless of separation)
is really hardly perceivable as a nonlocal interaction.
Rather we would tend to interpret such  effects as being a  part of the
laws of Nature, since such effects are forever everywhere the same. Such
an omnipresent effect is therefore effectively unobservable and we
would not in practice see any deviation from locality.

\section{Nonlocality can imply finetuning and multiple
point criticality}

We shall now argue that the assumption of nonlocality implies the realization
in Nature of what we sometimes
refer to as the  principle of multiple point criticality. For the purpose of
explaining why Nature seeks out the multiple point, we assume in accord with
the
argumentation of Section~3 that we have
fields $\phi$ depending on a single spacetime
point that interact nonlocally, in such a way that
the long distance remnants of the
nonlocal interactions between
fields $\phi(x)$ and $\phi(y)$  are the {\em same} for all pairs of
spacetime points
$x$ and $y$. As the reparameterization invariance of general relativity
implies this symmetry between spacetime points, we write our nonlocal
action as a nonlinear function of
reparameterization invariant integrals of the form

\begin{equation}
I_j \stackrel{def}{=} \int d^4x\sqrt{g(x)} {\cal L}_j(x)
\end{equation}
\noindent where the ${\cal L}_j$ denote the usual sort of terms in a local
Lagrangian density. An  ${\cal L}_j$ could, for example,  be a polynomial
of degree $n$ in
the (scalar) field $\phi(x)$: ${\cal L}_j=\phi^n(x)$ or the $k$th partial
derivative of such a field: ${\cal L}_j=\partial^k\phi^n(x)$ (somehow
made rotationally invariant).

We achieve nonlocality by considering actions
$S_{nl}( I_1,I_2,...,I_N)$
that are {\em nonlinear}
functions of the integrals $I_j$. Note that nonlinearity is tantamount to
nonlocality, because
nonlinearity in the quantities $I_j$ implies having integrals with more
than one integration variable; e.g., an action term $\propto I_iI_j$
is indeed nonlocal because

\[ I_iI_j=\int\int d^4xd^4y \sqrt{g(x)}\sqrt{g(y)}
{\cal L}(x)_i{\cal L}(y)_j\]

\noindent contains contributions from fields at independent
(and therefore in general different) spacetime points $x$ and $y$.
Note that  had we taken a linear function of the integrals $I_j$:
$S= \sum g_j I_j$,
we would get an ordinary local action.

An important property of the reparameterization invariant integrals $I_j$
is that any function of such integrals - even a nonlinear and thereby
nonlocal one - is also reparameterization invariant.
So we can say that we restrict the nonlocality allowed in our model
to the nonlocality that comes about, due to having
an action that is a nonlinear function of a
lot of integrals $I_j$ having integrands corresponding to the
various Lagrangian densities ${\cal L}_j$ being used.
Our speculation is that this form of nonlocality (formulated with the $I_i$'s)
is really the only form that can survive at long distances, when
reparameterization invariance is insisted upon (see however
Section~6 for
a generalization).

However, we now  want to argue that this restricted form of nonlocality
would not be easily observable and could therefore really exist
in Nature without ever having been observed as, for example, an ``action at
a distance'' sort of  nonlocality.
Rather we would say that the only traces of the restricted form of
nonlocality that we consider are
(some) solutions of finetuning problems.

Formally we can think of having the functional integral of Nature with
a nonlocal action $\hat{S}_{nl}$ that is a functional of the fields $\phi$
of the theory:

\begin{equation}
\int {\cal D}\phi e^{i\hat{S}_{nl}[\phi]}.
\end{equation}
where

\begin{equation}
\hat{S}_{nl}[\phi]\stackrel{def}{=}S_{nl}(I_1[\phi],I_2[\phi],\dots I_N[\phi])
\end{equation}
and $\phi$ is used as a symbol for all the fields of the theory.
As with any classical approximation for a field theory, it can be argued that
this functional integral is dominated by field values in the
neighbourhood of the field values $\phi_0$ for which the action is stationary:
\begin{equation}
\frac{\delta \hat{S}_{nl}}{\delta \phi}|_{\phi_0}=0 \label{eq2}. \end{equation}

\noindent Were the quantities $I_i$ effectively independent, we would deduce
from Eq.~(\ref{eq2})  that

\begin{equation}
\frac{\partial S_{nl}(\{I_i\})}{\partial I_j}=0\label{eq3}.
\end{equation}

If there are some necessary
relations between the $I_i$'s, because of their functional forms as
 functionals  of the same fields $\phi$, the $I_i$'s may be constrained to
take values in only some allowed region of the space spanned by the
$\{I_i\}$ (see Fig. 1 for an example).
In the event that $S_{nl}$ has an extremum on the border of the
allowed region, we should only require that the variation of $S_{nl}$
vanish along this border. In this event,

\begin{equation}
\frac{\partial S_{nl}(\{I_i\})}{\partial I_j}=\lambda a_j. \label{eq7}
\end{equation}
where the variation along the border obeys the restriction
$\sum a_jdI_j=0$ and $\lambda$ is a Lagrange multiplier. If the border is of
codimension greater than one, there will be a Lagrange multiplier for each
codimension and a corresponding contribution in Eq.~(\ref{eq7}).

We illustrate the idea of how a nonlocal reparameterization invariant action
can lead to finetuning by an example in  which we ignore
derivative terms in the action. Thus we consider only a nonlocal pure
scalar field potential type action, in which the
potential term is nonlocal at  very long distances in such a way that
the interaction is independent of the
separation of spacetime points. This is insured by taking a nonlocal
potential $\hat{V}_{nl}[\phi]=V_{nl}(\{I_j[\phi]\})$ that is a (nonlinear)
function of the quantities $\{I_j\}$.

We now seek the minimum
for a nonlocal potential $V_{nl}(\{I_j\})$ in a space spanned by
quantities $I_j/V$ - i.e., the volume densities of the quantities $I_j$.
For expositive purposes we consider the simple situation in which
there are just two quantities $I_1$ and $I_2$ where

\begin{equation}
I_1=\int d^4x\sqrt{g(x)}\phi(x)\stackrel{def}{=}\int d^4x
\sqrt{g(x)}v_1(\phi(x))
\end{equation}
and
\begin{equation}
I_2=\int d^4x\sqrt{g(x)}(\phi^4(x)-5\phi^2(x)+\phi(x))\stackrel{def}{=}
\int d^4x\sqrt{g(x)}v_2(\phi(x))
\end{equation}
where $v_2(\phi(x))$ is some polynomial that, for illustrative purposes,
is taken as being 4$th$
order; e.g., $v_2(\phi(x))=\phi^4(x)-5\phi^2(x)+\phi(x)$.

One should bear in mind that the integrals $I_1$ and $I_2$
of field polynomials over spacetime cannot take values
completely independent of each other. If, for instance,
the integral $I_2$ say of $v_2(\phi(x))$ over spacetime is required
to be rather
small, the value of $\phi$ cannot be too large over most of spacetime. This
in turn would limit the possible values of the
integral over spacetime of $\phi$ itself. Taking such relationships into
account leads to an allowed region of values for the $\{I_j\}$.
Including many polynomials in the fields $\phi$ can lead
to allowed regions that can be somewhat complicated. We shall continue to
restrict our example to the two quantities $I_1$ and $I_2$ defined above.
Figure 1 depicts schematically the allowed  region of $I_1$ and $I_2$ values,
with values of $I_1/V=\langle \phi \rangle$ plotted along the abscissa
and $I_2/V=\langle v_2(\phi) \rangle$ along the
ordinate. The average $\langle \rangle$ denotes an average over spacetime.
The part of the boundary of the convex envelope of allowed values
drawn as the heavy solid
curve corresponds to having a constant $\phi(x)$ in spacetime: $\phi(x)=
I_1/V$.

\begin{figure}
\vspace{12cm}

\caption{The nonlocal potential $V_{nl}(I_1,I_2)$ can have its minimum at any
point in the interior (e.g., point 3) or on the boundary (e.g., points 1 or 2)
of the convex closure
of allowed $(I_1/V,I_2/V)$ combinations (the cross-hatched area). The three
inserts show the characteristic form of the effective local potential
$V_{eff}$ at the three generic possibilities for the minima of $V_{nl}$. For
minima of $V_{nl}$ at interior points, $V_{eff}$ is just flat (see insert at
point 3). At minima of $V_{nl}$  on the heavy solid curve portion of the
boundary, the characteristic feature of $V_{eff}$ is one absolute minimum
(see insert at point 1) corresponding to $(I_1/V,I_2/V)$ combinations
realizable in a universe  with just one
(dominant) value of $\phi(x)$ in the vacuum. At minima of $V_{nl}$ located at
boundary points indicated by the heavy broken
line, the characteristic feature of $V_{eff}$ is {\em two} equally deep minima
(see insert at point 2), corresponding to $(I_1/V,I_2/V)$ combinations that
can be realized as the vacuum of a universe having different dominant
constant values of $\phi(x)$ in different spacetime subregions.}

\end{figure}

A priori, the nonlocal potential $V_{nl}(I_1,I_2)$ can
have its minimum at
any point in the interior (the
cross-hatched region of Figure 1) or on the boundary of the allowed region
(convex envelope). The  heavy solid curve of Figure 1
corresponds to the
$(I_1/V,I_2/V)$ combinations that can be realized in a universe
having just one dominant value of
(i.e., almost everywhere in spacetime constant)
$\phi(x)$ in the vacuum. Here the symbol $V$ denotes the
volume of the universe.
That is, $\phi(x)=\langle \phi \rangle$ for
almost all of space(time).
Allowed $(I_1/V,I_2/V)$ combinations, not lying on the
heavy solid curve portion of the boundary of the convex envelope,
cannot be realized in a universe
having a single dominant (for all spacetime)  constant value of $\phi(x)$.
However, such points can
be realized by means of a positively weighted linear combination
of points on the heavy solid curve.
Such points would correspond to a
universe the vacuum of which has different dominant constant values
of $\phi(x)$ in
different spacetime subregions, where the extent of these subregions is
proportional to the positive weights needed, in the combination
of the several constant values of $\phi(x)$, to
get a universe having the average values $\langle \phi \rangle = I_1/V$
and $\langle v_2(\phi) \rangle=I_2/V$.

In Figure 1, we also indicate with the points 1, 2 and 3 representatives for
the three generic classes of points, in the convex envelope of allowed
$(I_1,I_2)$
combinations, at which $V_{nl}(I_1,I_2)$ can have its minimum:
point 3 represents
the interior, point 1 represents the class of points on the heavy solid curve
coinciding with the boundary of the convex envelope,
and point 2 is a prototype for the remainder of the boundary of the convex
envelope.
It is reasonable
to claim that all of these 3 prototypes represent generic possibilities
- even though one might
a priori think that a minimum on the border would require some degree of
finetuning\footnote{A point seeking a minimum in the allowed
region would statistically often tend to accumulate somewhere along the
border.}.

A moment's
reflection can perhaps convince the reader that a point such as 3
can be obtained as a suitably (positively) weighted combination of infinitely
many points on the heavy solid curve in Figure 1. Points on this
heavy solid line in Figure 1 correspond to universes that can be realized
with fields that are almost everywhere equal to the average values of
these fields
(i.e., essentially the same constant value
for $\phi(x)$ at almost all spacetime points $x$):
$\forall x,\; \phi(x)=\langle \phi \rangle$.

A point on the portion of the solid curve coinciding with the
boundary of the convex envelope of allowed $(I_1/V, I_2/V)$
combinations - point 1 for example -
can only be obtained as a single-term
``combination'' of different
constant values of
$\phi(x)$ - namely the constant value of $\phi(x)$ at the point 1.

The final prototype point at which
$V_{nl}(I_1,I_2)$ can have its minimum - the interesting case as it
turns out - is  point 2, located on the boundary of the
convex envelope that is not on the heavy solid curve.
Such  a point corresponds to a universe unrealizable with a
\underline{single} constant (i.e.,
everywhere in spacetime  constant)
value of $\phi(x)$.

At such a point, there are only
two constant values of $\phi(x)$ (having one constant value
at  points in some spacetime subregion and the other constant value
at all other points in spacetime)
that together can participate in a weighted combination
that can realize the prototype  point 2.
These are the constant values, $\phi=\phi_A$ and
$\phi=\phi_B$,
at the points on the boundary of the convex envelope at which the heavy
broken line of universes, unrealizable with single constant values
of $\phi(x)$, is tangent
to the heavy solid curve corresponding to all universes that
are realizable with a single
value of $\phi(x)=\langle \phi \rangle$:

\begin{equation}
\frac{I_1}{V}=\langle \phi_{point\;2}\rangle =
w_A\phi_A+w_B\phi_B\;\;\;(w_A+w_B=1)
\end{equation}
where $w_A$ is proportional to the  extent of the spacetime region having the
constant value
$\phi_A$ and
$w_B$ to that of the spacetime region having the constant value $\phi_B$.

We want to examine the effective local potential in the three cases, in which
the nonlocal potential is located at the three types of points 1, 2 and 3.

The effective local potential $V_{eff}$ is defined
as that function of $\phi$ for which the
derivatives are equal to the corresponding (functional) derivatives of
the nonlocal potential $V_{nl}$.
We can think of $V_{eff}$ as the potential observed in a
laboratory  very small compared to the volume
of the universe and arbitrarily placed at some spacetime point.
The derivative of $V_{eff}$ is the change in $V_{eff}$
observed in the laboratory, when the value of the field $\phi$
is changed only in the laboratory and kept constant at all other points of
space.
If $\phi$ is changed by a finite amount in the laboratory,
the effective local potential can be integrated
up: $V_{eff}(\phi_a)-V_{eff}(\phi_b)=
\int^{\phi_b}_{\phi_a} V_{eff}^{\prime}(\phi)d\phi$.

Formally we make the definition

\begin{equation}
\frac{\partial V_{eff}(\phi(x))}{\partial \phi(x)} \stackrel{def}{=}
\frac{\delta V_{nl}(\{I_j[\phi]\})}{\delta \phi(x)}
\left|_{near\;\;min.} \right.=
\sum_i\left(\frac{\partial V_{nl}(\{I_j\})}{\partial I_i}\frac{\delta
I_i[\phi]}
{\delta \phi(x)}\right)\left|_{near\;\;min.} \right. \label{eq11}
\end{equation}
\[ =\sum_i\frac{\partial V_{nl}(\{I_j\})}{\partial I_i}
\left|_{near\;\;min.} \right. v_i^{\prime}(\phi(x))  \]
This definition implicitly assumes that, to a very good approximation,
$V_{nl}$ takes on its lowest
possible value. But this does not preclude small regions
of spacetime from having $\phi$ values that deviate, by essentially any
desired amount, from the average value(s) in the vacuum or vacua\footnote
{i.e., more than one vacuum in the,
for us, interesting case of competing vacua corresponding to different
phases in different regions of spacetime.}.
The subscript ``near min'' in this formula
denotes  the approximate ground state of the whole universe, up to deviations
of $\phi(x)$ from  its vacuum value (vacuum values for a multi-phase vacuum)
in relatively small regions.

As a solution to Eq.~(\ref{eq11}) we have

\begin{equation}
V_{eff}(\phi)=\sum_i \frac{\partial V_{nl}(\{I_j\})}{\partial I_i}v_i(\phi)
\label{eq12} \end{equation}
where the $v_i(\phi)$ are the (field polynomial) integrands
of the
``extensive'' (reparameterization invariant) quantities
$I_j=\int d^4x\sqrt{g(x)}v_j(\phi(x))$. We can identify the
$\frac{\partial V_{nl}(\{I_j\})}{\partial I_i}$ as intensive quantities
conjugate to the $I_i$.

That Eq.~(\ref{eq12})
solves Eq.~(\ref{eq11})  is easily seen by differentiating
Eq.~(\ref{eq12}) and using
that the right hand side of Eq.~(\ref{eq11}) is
$\sum_j \frac{\partial V_{nl}}{\partial I_j}\frac{\delta I_j[\phi]}{\delta
\phi}
=\sum_j \frac{\partial V_{nl}}{\partial I_j}v_j^{\prime}(\phi(x))$. The seeming
$x$-dependence of this right-hand side of Eq.~(\ref{eq11}) for  prescribed
values of $\phi(x)$  is effectively absent due, at the end,  to
the reparameterization invariance hidden in the form of the $I_j$'s.

We now proceed with a study of the effective potential $V_{eff}$
for the field configurations $\phi$ near
the minimum of $V_{nl}$, when this minimum is near one of the three types
of points 1, 2 and 3.

At an interior point of type 3, the absolute minimum of $V_{nl}$ is also
a local minimum and $\frac{\partial V_{nl}(\{I_j\})}{\partial I_i}=0$
for all $i$. Accordingly,

\begin{equation}
V_{eff}=\sum_i\frac{\partial V_{nl}}{\partial I_i}\left|_{near\;\; min.\;\;
at\;\;``3"} v_i(\phi) =0 \right. \end{equation}
So when $V_{nl}$ has its minimum in the interior, the effective potential
$V_{eff}$ is flat. Recall that an interior point such as 3
can be obtained as a suitably (positively) weighted combination of infinitely
many points on the boundary of the allowed region. This is related
to the fact that $V_{eff}$ has infinitely many
minima (because it is flat) at an interior point at which $V_{nl}$ has its
minimum.

If $V_{nl}$ has its minimum at a point of the type 1 or 2 (i.e., on the border
of the convex envelope), we have in general that

\begin{equation}
V_{eff}=\sum_i\frac{\partial V_{nl}}{\partial I_i}\left|_{near\;\; min.\;\;
at \;\;``1"\;\;or\;\;``2"} \;v_i(\phi) \neq0 \right. \end{equation}
because in general
$\frac{\partial V_{nl}(\{I_j\})}{\partial I_i}\neq 0$ at an
absolute minimum of $V_{nl}$ located on the boundary of the
convex envelope.

For the minimum of $V_{nl}$ located at a point of the type 1, there is
only one value of $\phi(x)$ realized in the vacuum (i.e., in extended regions
of spacetime) - namely the value $I_1/V$ at which the minimum
of $V_{nl}$ is located. This means that there is only one
value of $\phi$ assumed in extensive regions of the vacuum for the
nonlocal interaction and $V_{eff}$ has a {\em single} deepest minimum
 - namely at
$ \phi_1$ where the latter denotes the value of $I_1/V$ at the border
point 1 where $V_{nl}$ has its minimum.

That there is only one deepest minimum
is readily seen, by showing that the assumption of a second equally
deep minimum at some other value $\phi_C$ would lead to a contradiction.
First we make the observation that the gradient of $V_{nl}$, which
cannot be zero for a generic point of type 1, is perpendicular
to the tangent to the convex envelope at point 1. Secondly, note that the
line connecting $\phi_1$ with $\phi_C$ determines a chord of the
convex envelope that necessarily lies in the interior of the convex envelope.
A displacement away from point 1, along such a chord, has therefore always a
component along the gradient of $V_{nl}$. But moving along this chord,
defined by the  two equally deep minima in
$V_{eff}$ at respectively $\phi_1$ and $\phi_C$, corresponds to replacing
$\phi_1$ by $\phi_C$ (or vice versa) in a small spacetime region
at no cost in energy. This is inconsistent with the observation
that a displacement along this chord necessarily has a component
along the
gradient of $V_{nl}$. We conclude that $V_{eff}$ cannot have two
(or more) equally deep minima.

The most interesting case is that for which the minimum of $V_{nl}$ is located
at a point of type 2, with coordinates denoted as $(I_1,I_2)_{type\;2}$,
at a border point of the convex envelope of the
allowed region that can\underline{{\bf not}} be realized with
$(\phi(x),v_2(\phi(x)))=(I_1/V,I_2/V)_{type\;2}$.
It can be shown that, in order to realize
$(I_1/V,I_2/V)_{type\;2}$, only
the two constant contributions $(\phi_A,v_2(\phi_A))$ and
$(\phi_B,v_2(\phi_B))$  can participate
in the (unique) weighted combination.
A universe corresponding to the point \linebreak $(I_1/V,I_2/V)_{type\;2}$
could be realized with the field
\begin{equation}
\phi(x)=\left\{ \begin{array}{c} \phi_A \mbox{\footnotesize
{  for $x$ $\in R_A$ }}  \\
\phi_B \mbox{\footnotesize{ for $x$ $\in R_B$}}
                               \end{array} \right.. \end{equation}
where $R_A$ and $R_B$ are large regions of spacetime.
When a type 2 point is the location of the
minimum of $V_{nl}$, there are two and only two constant
$\phi$ values - namely $\phi_A$ and $\phi_B$ - which both are taken in
significant proportions in the vacuum of the universe.

It is interesting that two minima of $V_{eff}$ will be seen to have the
{\bf same} depth.
This is tantamount to finetuning, in
that the relation $V_{eff}(\phi_A)=V_{eff}(\phi_B)$ can be used to
eliminate a bare parameter (for example, the bare Higgs mass $m_H$).
Having two equally deep minima of $V_{eff}$, for $\phi_A$ and $\phi_B$,
is characteristic of a vacuum with two coexisting phases. This is
tantamount to being at the multiple point.

That we in fact have $V_{eff}(\phi_A)=V_{eff}(\phi_B)$, when the minimum
of $V_{nl}$ is at a type 2 point, can be seen by considering the directional
derivative of $V_{nl}$ along the line connecting the points
$(\phi_A,v_2(\phi_A))$ and $(\phi_B,v_2(\phi_B))$. This line is
parameterized by
\begin{equation}
(I_1/V , I_2/V ) = \xi (v_1(\phi_A),v_2(\phi_A)) +
(1-\xi) (v_1(\phi_B),v_2(\phi_B))
\end{equation}
with $\xi$ as the parameter. Along this line we have
\begin{equation}
\frac{dI_j/V}{d\xi} = v_j(\phi_A)-v_j(\phi_B)
\end{equation}
for $j=1,2$. The directional derivative is
\begin{equation}
\frac{dV_{nl}}{d\xi}= \sum_j \frac{\partial V_{nl}}{\partial I_j}
\frac{dI_j}{d\xi}
= \sum_j  \frac{\partial V_{nl}}{\partial I_j} ( v_j(\phi_A)-v_j(\phi_B))V
=V_{eff}(\phi_A) -V_{eff}(\phi_B) \label{AB}
\end{equation}
which means that, if Eq.~(\ref{AB})  is zero, the effective
potential $V_{eff}$ will take the same value in $\phi_A$ and
$\phi_B$. Let us emphasize that having demonstrated
$V_{eff}(\phi_A)=V_{eff}(\phi_B)$
amounts to having derived multiple point criticality
at least with finite
probability, i.e. in one generic situation.

\section{Examples of solving finetuning problems}

\subsection{Finestructure constants}
In presenting our finestructure constant
model, it is most straightforward to assume that there is a fundamental
truly existing regularization in the form of a lattice: the lattice
exists ontologically! Eventually we hope to replace the lattice assumption
with simply the assumption of some fundamental  regularization. We would
hope that any manifestation of the fundamental regularization would, in
the end, give us roughly the same predictions.

Having the lattice,  we can construct a lot of lattice-defined
extensive quantities meaning sums over the lattice of some functions
of the field variables (i.e., the link variables ) that  obey the symmetry
properties imposed. For example, when working with gauge theories,
gauge invariance is imposed on these  quantities. Such quantities
are the lattice equivalent  of the $I_j$'s in the abstract case.
A typical candidate
for such an $I_j$ would be the ``usual'' sum, over all plaquettes
in the lattice, of the Real part of the trace of the plaquette variable
in some representation $r$:
\begin{equation}
I_r = \sum_{\Box} Re(Tr(U_r(\Box))).
\end{equation}
With a nonlocal action $S_{nl}(I_1,...,I_N)$ depending on these
extensive quantities, one is likely to end up with several
coexisting vacua (phases) and an effective lattice gauge
theory, having an action constructed from the various $I_j$'s
that have coefficients (intensive quantities) that take values at the
multiple point. As a result, the free energy or rather
$logZ$ for these  various coexisting phases acquire equal values.

In principle, the next step is to use Monte Carlo computer simulation
to find the values of these coefficients to the $I_j$'s,
by studying the phase-diagram of the model chosen
- we would choose  our $SMG^3$ Yang-Mills theory on the lattice -
using the local action
\begin{equation}
S_{simulation} = \sum_j \beta_j I_j;
\end{equation}
i.e.,
in the space of all the coefficients $\beta_j$, one seeks all surfaces
at which there are singularities in derivatives of $\log Z$. The point
in $\beta_j$ space where
the maximum number of such surfaces meet is the multiple point.

Next, we compute the effective continuum coupling
for the lattice corresponding to the values of the $\beta_j$'s at the
multiple point ``corner'' of the phase
corresponding to the diagonal subgroup surviving the presumed Planck scale
spontaneous breakdown of the gauge group $SMG^3$.
These continuum couplings are then compared to the
experimental values of the finestructure constants.

Our results for the non-Abelian subgroups of the $SMG$ are in good
agreement with experiment,
provided that the $SU(2)$ and $SU(3)$ gauge subgroups of the $SMG$
are taken to be the diagonal subgroups of three isomorphic,
$SU(2)$ and $SU(3)$ respectively, subgroups contained in the underlying
gauge group  - i.e. in the $SMG^3$.
However, having comparable success in the case of $U(1)$ is contingent
upon taking into account certain details unique to $U(1)$:

\begin{itemize}
\item
Even in the continuum formulation it is possible to have terms
in the Lagrangian density of the form
$ cF_{\mu\nu}^{Peter}(x)F^{\mu\nu\;Paul}(x)$, where $F_{\mu\nu}^{Paul}$
and $F_{\mu\nu}^{Peter}$ are different Abelian gauge fields of $SMG^3$.
Having such terms, we can then also have $I$'s that are integrals
over such products of different Abelian gauge fields distinguished by the
labels ``$Peter$'' and ``$Paul$''.
Note that similar terms for non-Abelian
gauge theories would not be gauge invariant and therefore not allowed!

\item
For an Abelian group such as $U(1)^3$ as well as for the Abelian part of
$SMG^3$, all subgroups are invariant. As it is the invariant subgroups
of the gauge group that can be ``confined'' in the various ``phases'',
there
are a priori enormously many more confinement possibilities for Abelian
than for non-Abelian groups. It is therefore doubtful that it is realistic
to have
a multiple point at which all possible phases of an Abelian group come
together. Rather we should probably, in the case of an Abelian group,
envision a multiple point where a large number of phases convene.

\item
A suggested candidate for a multiple point is one aided by a Lagrangian
that has a high degree of symmetry. In this case, all phases that
transform into each other under this symmetry can be made to meet at
a multiple point, by choosing the parameters of the action in
such a way that any one of the symmetry-related phases convenes at the
multiple point.
As the symmetry that allows a lot of phases for the
$U(1)^3$ group contained in $SMG^3$ to meet at the multiple point, we
suggest a ``hexagonal''
symmetry of linear transformations that transforms the three Abelian
gauge fields into each other. We refer to this symmetry as ``hexagonal''
because, in a certain formulation, it is equivalent to
the discrete group of rotations that transforms  a three-dimensional
lattice of hexagonally packed sites - each having 12 nearest neighbours -
into itself.

To avoid confusion, it should be stressed that this lattice is in the covering
space of the $U(1)^3$ group and has nothing to do with the
latticification of space time that we also adhere to in the
present model, as one of probably many ways of implementing what we
assume to be the necessity of fundamental regularization.

\item
The concept of ``the'' diagonal subgroup of the group
$U(1)^3$ is not well-defined, in the way that it is in the case of  the
diagonal subgroup of say $SU(3)^3$. Therefore a bit of phenomenology or
guessing is
involved, in selecting the ``diagonal subgroup''  to be
identified with the weak hypercharge $U(1)$ gauge group found
in the Standard Model (this is a slightly weak point in our
model).

\item
In contrast to the non-Abelian case, the charges for the Abelian gauge group
can ultimately  only be defined in terms of the matter charge quantization
(in practice the fermions: quarks and/or leptons). It is only by
relating the compactified lattice gauge group to the
fermion (or matter) field charge quantization that a meaningful finestructure
constant can emerge from a study such as ours. But by adjusting
the global structure of the $SMG$ group\footnote{We define the standard
model {\em group} $(SMG)$ as the
factor  group  obtained  from  the  $SMG$  covering  group  $\br\times
SU(2)\times SU(3)$ by identifying the elements of the centre belonging
to      the      discrete      subgroup      $\{(2\pi,-\bunit^{2\times
2},e^{i\frac{2\pi}{3}}\bunit^{3\times3})^n|n \in \bz\}$:

\beq
SMG   \stackrel{def}{=}   S(U(2)\times   U(3))\stackrel{def}{=}
(\br\times   SU(2)\times
SU(3))/\{(2\pi,\bunit^{2\times   2},e^{i\frac{2\pi}{3}}\bunit^{3\times
3})^n|n \in \bz\}  \eeq

\nin The defining representation of $S(U(2)\times U(3))$  is  the
set of $5\times 5$ matrices

\beq     S(U(2)\times      U(3))_{def}\hat{=}      \left\{\left.\left(
\begin{array}{cc} \bu_2 & \begin{array}{ccc} 0 & 0 & 0 \\ 0 &  0  &  0
\end{array} \\ \begin{array}{cc} 0 & 0 \\ 0 & 0 \\ 0 & 0 \end{array} &
\bu_3 \end{array} \right ) \right| \begin{array}{l} \bu_2\in U(2),  \\
\bu_3\in U(3), \\ \mbox{det}\bu_2\cdot  \mbox{det}\bu_3=1  \end{array}
\right \} \eeq

\nin This (non-faithful) representation is  suggested  by  the
spectrum of the standard model.} to match the charge quantization
rule (Millikan-charge quantization with its sophisticated
extension to quarks), we believe that we get a reasonable normalization
method. It is therefore important for our scheme - in the Abelian case -
that we connect the charge quantization for fermions with
a specific compactified lattice gauge group model. It is the
multiple point for the latter that gives us our prediction for the Abelian
coupling of the fermions
used. In establishing the $U(1)$ normalization, we make use of
the approximation that, near the
multiple point, the fluctuations in the non-Abelian gauge group
lattice variables are so strong that we can, for the purpose of computing
the critical parameters for the Abelian part of the gauge group, assume
that the elements of $SMG$, belonging to both $U(1)$ and the non-Abelian
$SMG$ subgroups (i.e., $\bz_2\times \bz_3$),
can be  identified with the unit element. This amounts to taking for the
$U(1)$ degrees of freedom the {\em factor group} $U(1)/(\bz_2\times \bz_3)$.
\end{itemize}

In order to get the desired
phases  for the Abelian group $U(1)^3$ to convene at the multiple point
(according to  our up to now analytical and somewhat speculative
calculations), we have
to supplement our action with terms that are not
just the ones having  a simple continuum interpretation.
But since we want a large number of phases to come together at the multiple
point, the  requirements become somewhat
strict. So even though the lattice action is in principle arbitrary,
the action that emerges by requiring a large number of phase to convene
at the multiple point is in fact not so arbitrary.
In fact, the lattice action that we arrive at is less arbitrary
than for most other lattice
calculations.

In reality we have not to date simulated the desired
gauge group $SMG^3$ in all its glory, but rather attempted to use already
performed simulations for simpler groups $SU(2)$, $SU(3)$ and
$U(1)$. These are used, in conjunction with rather speculative analytical
calculations, to derive
the  continuum couplings corresponding to the diagonal subgroup, which
is purported to survive  after the
Higgs breakdown of $SMG^3$.
These calculations are not as accurate as would be desirable, but
seem to agree with the experimental values within their
estimated accuracy, which is of the order of $7\%$ to $10\%$
for the finestructure constants referred to the Planck scale
using the renormalization group. The most uncertain element in the
calculation turns out
to be the transition to the continuum limit. We want namely to use continuum
couplings in the relatively simple formula for
evaluating the diagonal subgroup gauge couplings from the couplings of the
factors in the cross product:
\begin{equation}
1/\alpha_{diag} = \sum_k 1/\alpha_k
\label{diag}
\end{equation}
where the finestructure constants for the latter are denoted
$\alpha_k$, while the (effective) finestructure constant for the
diagonal subgroup is denoted $\alpha_{diag}$.

For the case of the $U(1)$ gauge groups contained in the
$SMG^3$ group, we make the rather sophisticated hypothesis that
an interaction between the three $U(1)$ gauge groups
shows up ``in order'' to make possible a large number of phases,
corresponding to confinement of various subgroups of $U(1)^3$.
This interaction
is of the type that in the continuum language would
look like terms of the type $F_{\mu\nu\;Peter}F_{Paul}^{\mu\nu}$
in the Lagrangian density. Here $F_{\mu\nu\;Peter}$ and
 $F_{\mu\nu\;Paul}$ are the electromagnetic fields
for two different $U(1)$'s out of the three in $SMG^3$. Such
``interaction terms '' are possible for Abelian
gauge fields, while for non-Abelian gauge fields such terms
would not be gauge invariant. Because of such terms, the
formula Eq.~(\ref{diag}) for the coupling of the diagonal subgroup
fine structure constant is no longer even approximately
valid for the Abelian groups, for which the concept
of the diagonal subgroup is actually also not so well defined.
It turns out that we believe, for the Abelian diagonal
subgroup, that we get as a first approximation
$ 1/\alpha_{diag\;U(1)}=6/\alpha_{crit}$
for $U(1)^3$, rather than $1/\alpha_{diag\; SU(N)} = 3/\alpha_{crit\;SU(N)}$
as was the case  for the non-Abelian group 3-fold cartesian product
$SU(N)^3$ at the point where phases convene.

Our predictions, in the roughest approximation, are  that the
fine structure constants for the Lie algebras of the Standard Model
$U(1)$  and $SU(N)$ with $N=2,3$ deviate from
the critical values by being weaker, by respectively
a factor 6 and a factor 3.
Here we take the renormalisation
point at the fundamental scale - presumably the Planck scale
at which gravity may also have a phase transition.

Various refinements of the roughest approximation
involve a number of details,
such as using the multiple point in the plot of Bhanot\cite{bhanot}
rather than simply the critical couplings and
that the enhancement factor 6 gets corrected to 7 for
$U(1)^3$.

In Table \ref{t1} we compare our predictions -
i.e. the multiple point values as well as we could
compute them - with the experimental\cite{kim,amaldi} finestructure constants
transported, by the renormalization group in a desert
scenario, to the renormalization point taken to be the Planck
energy $\mu_{Planck}$=$ 1.2\cdot 10^{19}$ GeV.

\vspace{.4cm}

\begin{table}
\centering
\begin{tabular}{|c|cc||cccc|}\hline\hline
& $1/\alpha(M_Z) \stackrel{RG}{\longrightarrow}$ & $1/\alpha(\mu_{Pl})$ &
enh. fac. & $1/\alpha_{multi}$ & $1/\alpha_{pred}$ &
$1/\alpha_{Parisi}$ \\
\hline SU(3) & 8.47 $\pm$ 0.5 & 52.8 $\pm$ 0.7 & 3 & $18._{9}\pm 2$
& 56$\pm 6$ & $56._{7}$ \\ \hline
SU(2) & 29.7$\pm$ 0.2 & $48._8\pm 0.2$ & 3 & $16._5\pm 2 $
& $48._3\pm 6$ & $49._{5}$ \\ \hline
U(1) & $58.9 \pm 0.3$ & $54._3\pm 0.3$ & $6.8\pm 0.4 $ &$ 9.3\pm 0.5 $
&$ 63\pm 6 $ & 66 \\  \hline
\end{tabular}
\caption{Comparison of
theoretically predicted and experimental values of the three Standard Model
gauge couplings at the Planck scale. Experimental  values  (Delphi
results) have been extrapolated to the Planck scale using the renormalization
group with the minimal Standard Model. The uncertainties given
are only very crude estimates. The last column contains
$1/\alpha_{Parisi}\stackrel{def}{=}
\langle \frac{1}{N}Tr(U(\Box))\rangle\cdot 1/\alpha_{Naive\;MP}$
where $\alpha_{Naive\;MP}$ are  the naive continuum limit multiple
point ($MP$) couplings. The
$\alpha_{pred}$ are the predicted continuum
couplings based on Ref. 16 in the case of $U(1)$; in the
nonAbelian cases, the values given reflect our having done a crude
$U(1)$-related correction to the $\alpha_{Parisi}$.}
\label{t1} \end{table}
\subsection{Cosmological constant $\Lambda_{eff}=0$ at transition from
finite to infinite universe}

We have seen that our form of nonlocality admits
the generic possibility of a ground state having more than
one phase. This would amount to having multiple point criticality.
In terms of the multiple point criticality model for finetuning, one should
expect to find
the finetuned parameters observed in Nature at parameter values that
coincide with critical behaviour. Since the cosmological constant value
$\Lambda_{eff}=0$ corresponds to the border between finite and infinite
space spheres (universes), it is
not surprising that computer simulations\cite{ham,amb}
of quantum gravity indicate singular
behaviour at the value $\Lambda_{eff}=0$ for the
cosmological constant. According to our multiple point  model
for finetuning, the value of the cosmological constant realized in Nature is
indeed expected to be at a phase border. We can therefore claim that the
phenomenologically indicated value $\Lambda_{eff}=0$ coincides with our
theoretical prediction for the value of the cosmological constant.

\subsection{Our ``unfortunate '' prediction of the Higgs mass}

Likewise
we can claim that our finetuning model explains the physical
finetuning problem related to the hierarchy problem,
in the sense that Planck scale multiple point criticality implies that
multiple point parameters should also be at the border between Higgsed
and un-Higgsed phases.
At such boundaries there is a change of sign in $m_{Higgs}^2$ which, for
transitions that  are  weakly first order, would imply values of
$m_{Higgs}^2$ that, relative to the Planck scale, are strongly suppressed.
So in a certain sense we do indeed solve the physical problem
connected with the so called hierarchy problem which, incidentally,  has
often been used to argue for supersymmetry.
We have a scheme that, using one loop corrections, can
get logarithms into the condition for
the coexistence of phases, in such a way that the equation
imposing the coexistence of the Weinberg Salam Higgs phase and
the unbroken phase (in which $W^+$, $W^-$ and $Z^0$ would be
massless) becomes one involving the Higgs mass, or equivalently
the Higgs field expectation value, appearing only in a logarithm.
If this happens, we can argue that the finetuning mystery
of why the Higgs expectation value scale is so very low compared to say
the Planck scale is solved. Once the ratio is determined
by its \underline{logarithm} it can easily become very large.

The calculation of what happens when we impose the requirement
of the equality of the Higgs field potential depths for the
two minima corresponding to the mentioned phases - the  Higgs and
the unbroken (= Coulomb) phase - has in fact already been performed;
this specially
adjusted Higgs potential corresponds to the Linde-Weinberg
bound\cite{wein,linde}.
It is a slight modification relative to the (more well known)
Coleman-Weinberg bound\cite{cole}. In the Linde-Weinberg bound,
the Higgs bare mass is finetuned to the requirement
that the two minima be equally deep. This is precisely
what our prediction from nonlocality would suggest.
So we predict the Linde-Weinberg situation as the solution to the
problem of why the Higgs field expectation is  so small.
In this Linde-Weinberg situation, it is indeed such that
the vacuum expectation value of the Higgs field is
obtained from a logarithm and in this way comes to deviate exponentially from
the
input mass. In the philosophy of a fundamental scale, we
would of course take the input mass to be at the Planck scale (or whatever
the fundamental
scale is taken to be). This looks wonderful at first:
we have solved the problem of the small Higgs expectation value
by postulating fundamental nonlocality! The technical hierarchy problem,
that in going to different orders in perturbative calculations
include  quadratic divergences (if you do not have supersymmetry
at least) which are expected to be of the order of the cutoff scale,
is now talked away by saying that in going to different orders
in the perturbative calculations, we have for each order to recalculate
the amount of spacetime volume which is in the unbroken phase
(and the amount which is in the Higgs phase). After this recalculation- which
is not really
done of course since we do not actually know the nonlocal action-
we find that we have just to take the Linde-Weinberg case and,
in this way, the Higgs mass and expectation value are not truly
renormalized away.

However, a bit worryingly for our prediction, the Linde-Weinberg
bound does not agree so well with experiment: it predicts
(and this is accordingly also our prediction from nonlocality) a mass
which, using the usual expectation value known from
experimental weak interactions, turns out to be 7.8 GeV.
This is a failure of our model, but in a somewhat feeble attempt
to rescue it, one could claim that there could be several Higgs fields,
in which case the simple Linde-Weinberg calculation would not hold true.
Still one could hope perhaps to retain the exponential
behaviour of the Higgs field expectation value and
thereby still
solve the hierarchy-related problem of why the Higgs particle is so light.
But the necessity for this kind of rescuing procedure is of course really the
way an incorrect theory is revealed: more and more
``crutches'' are needed to make it function.

\subsection{Mass Hierarchies of Quark-Lepton Generations}
As one of the finetuning mysteries we also count the question of why
the masses for most quarks and leptons are so small compared to the
weak interaction scale, which is the mass you would expect
if the Yukawa couplings were simply of order unity \cite{brijcdf}.
The explanation suggested is that, in the physics at the
fundamental scale, there are some Higgs fields that get
relatively small expectation values, much in analogy to
the above speculation about the Weinberg-Salam Higgs field. They might
even easily be exponentially light, because of
their masses being determined via a logarithm. But
even just the presence of the phase transition makes the Higgs masses small.
After all, it is on the border of positive and negative mass
squared that we have the separation line between the Higgs and the
Coulomb phases. This then means that the transitions between
left and right handed components of fermions, which
need such Higgses for their occurrence, are suppressed.
In some recent work \cite{froggatt}, we have explored
 the mass hierarchy of fermions
(leptons and quarks)
as a consequence of Higgsed
gauge symmetries that are only weakly broken,
due to the Higgs fields having only ``small'' expectation
values.
Such a mechanism can rather naturally be the explanation for the
large gaps between the generations in the mass spectrum.

{}From the point of view of a very general picture of approximately conserved
quantum numbers, we have searched for clues as to which approximately
conserved gauge quantum numbers should exist beyond the Standard Model.
In particular we have studied the extra gauge quantum numbers in the
$SMG^3$ model, so popular with us to explain the values of the fine
structure constants. This model can naturally explain the generation mass gaps,
but there is a problem to get the top quark mass sufficiently heavy
compared to the bottom quark and tau lepton masses. However adding yet
another extra Abelian gauge group $U(1)_f$ helps.


\subsection{Strong CP-conservation also at a meeting of phases}
Yet another finetuning problem is the strong CP-problem.
Schierholz has recently studied the $\Theta_{QCD}$ dependence of the
first order deconfining phase transition\cite{schierholz}. He finds
that in the continuum limit there is a critical point at $\Theta_{QCD} = 0$,
where the confinement phase corresponding to $\Theta_{QCD} = 0$ meets the
``Higgs'' phase corresponding to $\Theta_{QCD} \neq 0$. It then follows,
assuming that QCD is in the confinement phase, that $\Theta_{QCD} = 0$ and
CP is conserved by the strong interactions.
                             However, from our point
of view, we look for a more ``ontological''
type of solution and do  not accept that Nature at the bare level
should be precisely renormalized to, for example, reveal the confining phase
of long distance Q.C.D. We can nevertheless,
in the spirit of our multiple point principle above,
use the phase diagram of Schierholz to suggest
the possibility that $\Theta_{QCD}=0$ can be characterized
as a meeting point for phases. Therefore even the strong
CP-problem of why  $\Theta_{QCD}$ is so small can find
an explanation derivable from the coexistence of phases,
in the same spirit as our solution of
the other finetuning problems.

\section{Two-position fields/particles replicated at one position}

In Sections~3 and 4 we  have suggested
that, {\em if fields are
defined the usual way as functions of spacetime points} (and if we for
simplicity ignore
short distance nonlocalities),  we can interpret
long distance nonlocality that is {\em independent of spacetime}
as being incorporated into the laws of Nature rather
than being observable in some offensive way.
However, having once relinquished the principle of locality in this mild
ontological sense, there is no longer any compelling reason to assume that
fields depend on just one spacetime point!
Rather it becomes quite natural to contemplate
the possibility of having, for example,  a field $\phi(x,y)$ that depends
on {\em two} spacetime points $x=x_{\mu}$ and $y=y_{\mu}$. If we
for simplicity take $\phi(x,y)$ as a scalar field, it transforms under
reparameterization transformations (i.e., diffeomorphisms),
in both $x$ and $y$:
\begin{equation}
\phi \rightarrow \phi^{\prime} \;\;\mbox{where}\;\;\;
\phi^{\prime}(x^{\prime},y^{\prime})= \phi(x(x^{\prime}),y(y^{\prime})).
\end{equation}
Physically a field such as $\phi$ is just a function of
a couple of spacetime points regarded as abstractly defined (i.e.
coordinate independent) events.

The integrals that can be used for constructing a nonlocal
but still reparameterization invariant action, depending on such
double-position fields $\phi(x,y)$, can hardly be imagined to  be anything but
double integrals of the form
\begin{equation}
\int d^4x \int d^4y {\cal F}(\phi(x,y), \phi(x), \phi(y), \partial \phi(x,y)/
\partial x^{\mu}, ...).
\end{equation}

Roughly such a model can be thought of as one in which spacetime is
8-dimensional rather than just 4-dimensional but in which there are two
types of particles:

a) ``ordinary'' particles ( or fields ) having only one position
and really only depending on four out of the eight coordinates,
e.g. on $x$ but not on $y$.

\noindent and

b) ``double position particles (or fields)'' (e.g., $\phi(x,y)$)  that can
take values in the entire 8-space.

In practice we presumably have something
we may call ``vacuum'' for both sorts of fields: vacuum values in the
classical approximation are
constant (zero say)  over most combinations of $x$ and $y$. If ``we'' now
are primarily composed of ``ordinary'' particles, we can not readily
interact with the Fourier components of the $\phi(x,y)$ field
unless these components  have zero momentum along either $x$
or $y$.
Genuine excitations of $\phi(x,y)$ locally {\em in the 8-dimensional space}
can only occur by interaction of {\em two} ordinary particles and therefore
are presumably rather suppressed.
This makes it very difficult in practice to
discover the nonlocality related to the two-position fields,
{\em unless} there are some huge amounts of
matter in 8-space so to speak.

If reparameterization
invariance is not to be broken spontaneously, we must have
fields - for example $\phi(x,y)$ -that are constant in all 8-space points
$(x,y)$  except
along the diagonal $x=y$ (and, presumably, infinitesimally close to $x=y$).
However, the absence of nonlocality that we experience phenomenologically
is probably insured if there is not spontaneous breakdown under
reparameterizations in a {\em local} region. So if the reparameterization
in one neighbourhood differed from that prescribed by reparameterization
invariance at another very
far removed region, this might well not be observed as a breaking
of locality.
So a priori it would not be forbidden phenomenologically if
the field $\phi(x,y)$ takes on  some other values (i.e., departing
{}from the almost everywhere dominant constant vacuum value) along a thin band
representing a graph of a function yielding $y$ as function of $x$.
We assume that the structure along this  band is the same all over,
so that there is still reparameterization invariance under
the special type of transformation that transforms  the $x$ and $y$ at a point
$(x,y)$ in the band in the same way.
Field configurations corresponding to this band make up  a 4-dimensional
manifold in the
8-dimensional $(x,y)$ space, along which there can be a systematic
communication between a spacetime point $x$ and its image $y$.

If we had  efficient communication by nonlocality between say
$x$ and $y$ due to the above-outlined spontaneous breakdown of
reparameterization invariance, one can enquire as to whether
this effect would be {\em perceived} as a breaking of locality.
Probably not:  rather we would {\em interpret
the related space time points} - the ones with $(x,y)$ on the band -
as representing different degrees of freedom at a
single space time point $x$ say. Because we would experience
the spacetime point
$x$ and its image $y$  as the same spacetime point, locality
is effectively restored. Concurrent with this, we would experience
a replication of the field degrees of freedom at one spacetime point!
In Nature we seem to see a 3-fold replication with
respect to the fermions, in the sense that we observe
three generations. A tripling of the number of fields can
easily be achieved with
(just) two-position fields that are applied a couple of times:
the two-position field may have several ``bands'' as proposed, so that
one point - $x$ say - can be connected to several (e.g. two) far away points
$y$, by two different bands in the same $\phi(x,y)$-field.

Indeed, in  the experimentally supported Standard Model, we find a trinity
of similar (but not exactly replicated) field types:
the three generations of quarks and leptons!
But the three generations found
in Nature are not exact replicas of each other, as one at first might expect
if these truly represented
particles of the same sort just at different points in space and time.
However these generations may correspond to superpositions of
states at different related spacetime points;
there is also the possibility of
some sort of (later) spontaneous breakdown of the symmetry
between the different related space time points. There
is therefore no necessity  for  perfect symmetry between the
different generations, in order to uphold the interpretation that these
are due to the nonlocality with spontaneously broken reparameterization
symmetry.

However, we would expect that a 3-fold (approximate) replication mechanism
due to nonlocality would not only triple the quark and
lepton fields but also the boson fields! In fact, such a tripling of boson
fields
is an intrinsic feature of our long standing
$SMG^3$ ``anti-grand unification'' gauge group model. In this model,
we predict that  the values of the finestructure constants at
the multiple point
of the phase diagram for the gauge group $SMG^3$ should agree with
experimental values.
For the moment let us content ourselves with the observation
that nonlocality can easily lead to a picture in which
not only the fermions are tripled but also, essentially unavoidably, the
bosons. That is to say, we would predict,
roughly speaking, 3 photon-types, three $W^+$'s, three $W^-$ 's, three
$Z^0$'s, and 24 gluons.
We may also need to give large masses to some predicted but
not observed gauge bosons.
Presumably there should even be more Higgses or replacements for them.

\section{Conclusion}

Relinquishing  strict adherence to the principle of locality has
several attractive features from a theoretical point of view.
In the spirit
of Random Dynamics\cite{frogniel,randyn}, we have also investigated
the possibility of removing this principle from the list of initial
assumptions necessary for constructing a fundamental theory.

Starting with the problem of the cosmological constant
being almost zero - from a Planck scale point of view at least
(nowadays it seems from cosmological studies that it may not be
exactly zero) - we have argued that there should be a  breakdown of the
principle of locality for field interactions from a
fundamental point of view.  Actually, a  theory with nonlocality
that retains reparameterization invariance provides a
promising approach to explaining  a series of finetuning problems.
Essentially
all of the well known finetuning mysteries in high energy physics are
solved: 1) the vanishing of the dressed cosmological constant; 2) the small
Higgs field expectation values and masses; 3) the hierarchy of quark and
lepton masses; 4) strong CP conservation and 5) the
multiple point values of finestructure constants.

It is  possible to interpret our multiple point
criticality approach
for predicting
      gauge  finestructure constants in terms of nonlocal interactions,
because these latter essentially imply such  a principle.
These finestructure constant predictions agree very well
indeed - within something like 7 to 10 percent - with the three
experimentally determined finestructure constants
of the Standard Model, when
the latter are extrapolated to the Planck scale using a desert scenario.

We think that the principle of multiple point criticality - which
essentially asserts the coexistence of a number of phases -
leads to such an impressive number of good
predictions for finetuned quantities that one is almost
forced to take it seriously! One detail is however a bit disappointing:
we predict that Nature has the Linde-Weinberg Higgs mass of about 8 GeV
and, in the same connection, a top-quark mass of less
than about 90 GeV in order to have the Linde-Weinberg situation at all.

Having once renounced a strict principle of locality at the fundamental scale,
it is possible for fields to depend on
more than one spacetime point (separated by large distances).
We propose that such fields might cause a spontaneous breakdown of
reparameterization  invariance, so that distant points in spacetime become
related.
Degrees of freedom at distant points, related
by this breakdown,
would be interpreted as several degrees of freedom at the same point.
Such a field replication mechanism that comes from ``explaining away''
ontological nonlocality would be welcome, as a possible explanation
for the 3-fold replication seen in the three generations of quarks and
leptons. That a 3-fold replication mechanism for fermions would,
probably unavoidably, also provide a 3-fold replication of bosons is
also  a very welcome prediction in the context of our long standing
``anti-grand unified'' model that uses the gauge group $SMG^3$ (i.e., the
3-fold Cartesian product of the Standard Model Group). This gauge group,
the Planck scale breakdown of which yields the normal $SMG$ in our model,
is an important ingredient in our predictions of gauge couplings using
multiple point criticality.

\section{Acknowledgements}
We send thanks to  L. Laperashvili, G. Lowe, K. Olsen, S.E. Rugh,
D. Smith and
C. Surlykke for useful and stimulating discussions. Financial support
{}from INTAS Grant 93-3316 and EF Contract SC1 0340 (TSTS) is gratefully
acknowledged.

\end{document}